# The *u*-series: A separable decomposition for electrostatics computation with improved accuracy

Cristian Predescu,[1,†] Adam K. Lerer,[1] Ross A. Lippert,[1] Brian Towles,[1] J.P. Grossman,[1] Robert M. Dirks,[1,*] and David E. Shaw[1,2,†]

[1] D. E. Shaw Research, New York, NY 10036, USA.

[2] Department of Biochemistry and Molecular Biophysics, Columbia University, New York, NY 10032, USA.

† To whom correspondence should be addressed.

Cristian Predescu

    E-mail:    Cristian.Predescu@DEShawResearch.com

    Phone:    (212) 478-0433

    Fax:    (212) 845-1433

David E. Shaw

    E-mail:    David.Shaw@DEShawResearch.com

    Phone:    (212) 478-0260

    Fax:    (212) 845-1286

* Deceased February 3, 2015

## Abstract

The evaluation of electrostatic energy for a set of point charges in a periodic lattice is a computationally expensive part of molecular dynamics simulations (and other applications) because of the long-range nature of the Coulomb interaction. A standard approach is to decompose the Coulomb potential into a near part, typically evaluated by direct summation up to a cutoff radius, and a far part, typically evaluated in Fourier space. In practice, all decomposition approaches involve approximations—such as cutting off the near-part direct sum—but it may be possible to find new decompositions with improved tradeoffs between accuracy and performance. Here we present the *u-series*, a new decomposition of the Coulomb potential that is more accurate than the standard (Ewald) decomposition for a given amount of computational effort, and achieves the same accuracy as the Ewald decomposition with approximately half the computational effort. These improvements, which we demonstrate numerically using a lipid membrane system, arise because the *u*-series is smooth on the entire real axis and exact up to the cutoff radius. Additional performance improvements over the Ewald decomposition may be possible in certain situations because the far part of the *u*-series is a sum of Gaussians, and can thus be evaluated using algorithms that require a separable convolution kernel; we describe one such algorithm that reduces communication latency at the expense of communication bandwidth and computation, a tradeoff that may be advantageous on modern massively parallel supercomputers.

# I. Introduction

Evaluation of the electrostatic energy for a set of point charges in a periodic lattice is required for a number of applications, including molecular dynamics (MD) simulations, electronic structure calculations, crystallographic calculations, and hydrodynamic simulations. The long-range nature of the Coulomb interaction leads to well-known computational challenges arising from the inefficiency of the direct summation of all pairwise interactions.[1–3] For MD simulations of biophysical systems in aqueous solution with explicitly modeled solvent[4]—the application that motivates us here—the unscreened Coulomb interactions are non-negligible even at distances as large as 100 Å, and thus extend over the entirety of most simulated systems. Furthermore, when using spatial-domain decompositions[5] on massively parallel computers, the long range of the Coulomb interaction leads to communication latency that limits parallel scaling.[6,7]

Many efficient approaches for evaluating the total electrostatic energy are based on Ewald's decomposition,[2,8] which divides the Coulomb potential function $1/r$ into short-range and long-range parts, also referred to as the "near" and "far" parts, respectively. The near part is effectively range-limited, and is computed by direct summation up to a cutoff radius $r_c$. The far part is typically computed as a sum in $k$-space (also known as reciprocal or Fourier space), where it too is effectively range limited, and can be efficiently computed by methods such as optimized Ewald summation,[9,10] the particle-mesh Ewald algorithm (PME),[11,12] the particle-particle particle-mesh method,[13] the fast Fourier Poisson method,[14] or Gaussian split Ewald (GSE).[15]

The near interactions in the Ewald decomposition decay rapidly but do not vanish at $r_c$, leading to various undesirable artifacts depending on the specific implementation. The typical approach of abruptly setting the near forces to zero beyond $r_c$ is equivalent to truncating the near electrostatic energy and shifting it downwards so that it vanishes continuously at the cutoff. This is known as the "cut-and-shift" approach,[16] and it introduces a small, constant error in the electrostatic energy at distances up to $r_c$. More sophisticated techniques based on switching functions have been applied to the Ewald near term to increase its smoothness at $r_c$; such techniques eliminate the error in the electrostatic energy up to the distance at which the switching function takes effect, thus localizing the error near $r_c$, but have not been found to measurably reduce errors in actual biomolecular simulations.[17] There has also been some work exploring decompositions based on screening charge distributions,[18–20] which can be made continuous at $r_c$, but which still exhibit errors at distances less than $r_c$.

In the present work, we introduce the *u-series*, a new decomposition of the Coulomb potential into near and far parts that is constructed to be exact up to the cutoff and continuous at the cutoff. The far part of the *u*-series is essentially the long-range portion of a bilateral series (a series whose summation index runs from −∞ to ∞) presented by Beylkin and Monzón,[21,22] who observed that the function $1/r$ can be well approximated with uniform relative error by a bilateral series of suitably scaled Gaussians. (The name "*u*-series" was chosen to reflect this property of *uniform* relative error). The near part is $1/r$ minus the far part, truncated at $r_c$ (hence the near and far parts sum to exactly $1/r$ at distances up to $r_c$). We construct the decomposition in such a way that the near part vanishes and is continuously differentiable at $r_c$, without recourse to approximations that perturb the total energy (e.g., shifting or switching). As a result, the magnitude of the Coulomb interaction error near $r_c$ is significantly decreased relative to the Ewald decomposition.

Our approach was motivated by the hypothesis that moving the electrostatic energy error from before $r_c$ (as in the Ewald decomposition) to beyond $r_c$ would improve overall accuracy due to self-cancellation of long-range errors in a charge-neutral system. This intuition is borne out empirically by numerical experiments in which we find a substantial improvement in the accuracy of simulation observables for a lipid membrane system. Alternately, the *u*-series can match the accuracy of the Ewald decomposition with reduced computational effort; based on our numerical tests, we estimate that *u*-series needs only about half the computational effort of Ewald for the same accuracy. In practice, these computational savings are effected by using a smaller cutoff radius for the near part and/or a coarser grid when employing a mesh-based method to evaluate the far part.

In certain situations, the *u*-series may have certain additional advantages: Its far part is a sum of Gaussians, which are separable functions (that is, products of one-dimensional functions; e.g., $g(x,y,z) = f(x)f(y)f(z)$), allowing its far part—unlike the far part of the Ewald decomposition—to be evaluated by algorithms that take advantage of this separability. Three-dimensional convolutions of separable kernels are computationally less expensive than convolutions of non-separable kernels,[23] and we will show that they also admit implementations with lower communication latency on parallel machines. In particular, we present a real-space algorithm that evaluates the far part with a lower communication latency than standard Fourier space methods, at the expense of some overhead in communication bandwidth and computation. Such a tradeoff may be beneficial on massively parallel supercomputers, providing an additional potential performance advantage for the *u*-series beyond that which is illustrated in our numerical experiments.

## II.     The Problem of Evaluating Electrostatic Interactions

We begin by reviewing the problem of evaluating the electrostatic energy for periodic systems. We briefly discuss Ewald's method and refer the reader to refs. 1–3 for additional details. For an in-depth treatment of the structure of the electrostatic potential for periodic charge distributions, we refer the reader to the original analysis of de Leeuw et al.[24]

Let $\rho(\mathbf{r})$ be a charge density that satisfies periodic boundary conditions $\rho(\mathbf{r}) = \rho(\mathbf{r} + \mathbf{n})$, where $\mathbf{n}$ denotes a vector $(m_x L_x, m_y L_y, m_z L_z)$ with $m_x, m_y, m_z$ integers, and $L_x, L_y, L_z$ the side lengths of an orthorhombic unit cell of volume $V = L_x L_y L_z$. The unit cell, denoted by $\Omega$, is assumed to be charge neutral (that is, $\int_\Omega \rho(\mathbf{r}) d\mathbf{r} = 0$). The charge density $\rho(\mathbf{r})$ can be continuous, discrete, or a combination of the form

$$\rho(\mathbf{r}) = \rho_c(\mathbf{r}) + \sum_\mathbf{n} \sum_{i=1}^{N} q_i \, \delta(\mathbf{r} - \mathbf{r}_i - \mathbf{n}), \qquad (1)$$

where $\rho_c(\mathbf{r})$ is the continuous part of the distribution, $q_i$ denotes the magnitude of the point charge at position $\mathbf{r}_i$ within the unit cell, and $\delta$ is the Dirac delta function. The electrostatic potential $\Phi(\mathbf{r})$ generated by the infinite, periodic distribution of charge is formally given by

$$\Phi(\mathbf{r}) = \int_{\mathbb{R}^3} \frac{\rho(\mathbf{r}')}{|\mathbf{r}-\mathbf{r}'|} d\mathbf{r}'. \qquad (2)$$

The potential defines the electrostatic energy $\mathcal{E}$ of the unit cell by the formula

$$\mathcal{E}(\rho) = \frac{1}{2} \int_\Omega' \rho(\mathbf{r}) \Phi(\mathbf{r}) d\mathbf{r}. \qquad (3)$$

As is customary, the prime in the integral of Eq. (3) means that the potential has the Coulomb singularity at $\mathbf{r}_i$ removed when interacting with the point charge $q_i$ (which itself was the source of the singularity). It can be shown that $\Phi(\mathbf{r})$ can be expressed as

$$\Phi(\mathbf{r}) = \sum_{\mathbf{k} \neq 0} 4\pi C_\mathbf{k} e^{-i\mathbf{k}\cdot\mathbf{r}} / |\mathbf{k}|^2, \qquad (4)$$

where **k** runs over what are referred to as vectors of the reciprocal lattice, $\mathbf{k} = (2\pi m_x / L_x, 2\pi m_y / L_y, 2\pi m_z / L_z)$, with $m_x, m_y, m_z$ integers, and

$$C_\mathbf{k} \triangleq \frac{1}{V} \int_\Omega \rho(\mathbf{r}) e^{i\mathbf{k}\cdot\mathbf{r}} d\mathbf{r}.$$

By the Fourier inversion theorem, the series

$$\rho(\mathbf{r}) = \sum_\mathbf{k} C_\mathbf{k} e^{-i\mathbf{k}\cdot\mathbf{r}} \qquad (5)$$

recovers the periodic charge density.

It is not straightforward to evaluate the electrostatic energy from Eq. (4) because of the need to remove the self-interaction energy. Ewald provides a resolution to this problem by dividing the Coulomb potential into near and far parts:[2,8]

$$\frac{1}{r} = \frac{1}{r}\operatorname{erfc}\left(\frac{r}{\sqrt{2}\sigma}\right) + \frac{1}{r}\operatorname{erf}\left(\frac{r}{\sqrt{2}\sigma}\right), \qquad (6)$$

where $\sigma > 0$ is an arbitrary length scale. The splitting can be physically rationalized[18,19] by canceling the charge distribution $\rho(\mathbf{r})$ in Eq. (2) with a smoother version of itself obtained by convolution with a Gaussian of width (standard deviation) $\sigma$. The second term in Eq. (6) (the far part) then represents the electrostatic potential generated by a unit charge that is distributed as a Gaussian of width $\sigma$ centered about the origin. It can be verified that

$$\frac{1}{r}\operatorname{erf}\left(\frac{r}{\sqrt{2}\sigma}\right) = \frac{1}{(2\pi\sigma^2)^{1/2}} \int_1^\infty \frac{1}{u^{3/2}} \exp\left(-\frac{r^2}{2\sigma^2 u}\right) du, \qquad (7)$$

which shows that the far part in the Ewald decomposition is a continuous superposition of Gaussians of width at least $\sigma$. Similarly, the near part is a superposition of Gaussians of width at most $\sigma$.

The near and far parts of the Ewald decomposition are substituted for the Coulomb interaction in Eq. (2), resulting in the near and far potentials, respectively. The integral involving the near potential converges absolutely in real space, where it is truncated at a suitable cutoff radius $r_c$ and computed as an explicit sum of pairwise interactions.

The far part of the electrostatic potential,

$$\mathcal{F}^{\sigma}_{\text{ewald}}(r) \triangleq \frac{1}{r}\text{erf}\left(\frac{r}{\sqrt{2}\sigma}\right), \tag{8}$$

generates an absolutely convergent sum in $k$-space, which takes the form

$$\begin{aligned}\Phi^{\sigma}_{\text{far}}(\mathbf{r}) &\triangleq \int_{\mathbb{R}^3} \rho(\mathbf{r}') \mathcal{F}^{\sigma}_{\text{ewald}}(|\mathbf{r}-\mathbf{r}'|)d\mathbf{r}' \\ &= \sum_{\mathbf{k}\neq 0} \frac{4\pi}{|\mathbf{k}|^2} e^{-\frac{\sigma^2|\mathbf{k}|^2}{2}} C_{\mathbf{k}} e^{-i\mathbf{k}\cdot\mathbf{r}}.\end{aligned} \tag{9}$$

The far part of the electrostatic energy is then defined by

$$\begin{aligned}\mathcal{E}^{\sigma}_{\text{far}}(\rho) &\triangleq \frac{1}{2}\int_{\Omega} \rho(\mathbf{r}) \Phi^{\sigma}_{\text{far}}(\mathbf{r})d\mathbf{r} - \frac{1}{\sqrt{2\pi\sigma^2}}\sum_{i=1}^{N} q_i^2 \\ &= V\sum_{\mathbf{k}\neq 0} \frac{2\pi}{|\mathbf{k}|^2}|C_{\mathbf{k}}|^2 e^{-\sigma^2|\mathbf{k}|^2/2} - \frac{1}{\sqrt{2\pi\sigma^2}}\sum_{i=1}^{N} q_i^2,\end{aligned} \tag{10}$$

where the constant $\sum q_i^2$ term subtracted in Eq. (10) is the self-interaction energy.

The cost of computing the near potential is proportional to the number of particles and to the volume of interaction $4\pi r_c^3 / 3$ for each particle. The cost of evaluating the $k$-space sum depends on the specific method used. For direct Ewald summation, a *k*-space cutoff $k_c$ is set to a fixed multiple of $1/\sigma$, and the number of *k*-space lattice points in the ball of radius $k_c$ is proportional to the volume $4\pi k_c^3/3$. For grid-based techniques such as PME and GSE, the mesh spacing of the grid is proportional to $\sigma$ (for a given amount of accuracy), and computational effort scales at least linearly with the number of grid points. For both methods, the cost of computing the far potential is thus at least proportional to $1/\sigma^3$.

## III. The *u*-series

In this section we introduce the *u*-series, a new decomposition of the Coulomb potential for Ewald-like electrostatics computations. Our starting point is a previously known[21,22] bilateral series approximation for $1/r$, which we study in Section IIIA. In Section IIIB we show how this bilateral series approximation leads directly to a decomposition of $1/r$. In Section IIIC we make small but important modifications to this decomposition to improve its accuracy and to make it more suitable for use in MD simulations, resulting in the decomposition we call the *u*-series.

### A. *The bilateral series approximation for 1 / r*

The function $1/r$ can be approximated by a bilateral infinite series of Gaussians of geometrically spaced widths, which we refer to as the *bilateral series approximation* (BSA).

Let $G_\sigma(r) = e^{-r^2/2\sigma^2}/\sqrt{2\pi\sigma^2}$ be a Gaussian of width $\sigma$, and let $b > 1$ be a positive constant (the *base*). The bilateral series approximation for $1/r$ is defined as

$$\mathcal{B}_b^\sigma(r) \triangleq 2\ln(b) \sum_{j=-\infty}^{\infty} b^{-j} G_\sigma(b^{-j}r) = \frac{2\ln(b)}{\sqrt{2\pi\sigma^2}} \sum_{j=-\infty}^{\infty} \frac{1}{b^j} \exp\left[-\frac{1}{2}\left(\frac{r}{b^j\sigma}\right)^2\right]. \tag{11}$$

The BSA is motivated by the identity

$$\int_0^\infty G_\sigma(rt)dt = \frac{1}{r}\int_0^\infty G_\sigma(u)du, \quad (12)$$

which follows from the change of variables $u = rt$. Rearranged, this identity provides a well-known integral expression for $1/r$,

$$\frac{1}{r} = 2\int_0^\infty G_\sigma(rt)dt. \quad (13)$$

A quadrature of Eq. (13) with geometrically spaced quadrature points (i.e., $t = b^{-x}$) leads directly to the BSA approximation for $1/r$ in Eq. (11).

We are not the first to observe that $1/r$ may be represented by the series of geometrically scaled Gaussian terms given in Eq. (11). Notably, Beylkin and Monzon[21,22] have previously made use of the same observation as a starting point to develop approximations to $1/r$ with bounds on the relative error for a finite range $\delta \le r \le 1$, where $\delta > 0$. Such approximations have been used in the context of electronic structure calculations,[25] and other approximations that use a sum of Gaussians to represent an electrostatic potential have found use in that context[26] or in quantum mechanics/molecular mechanics calculations.[26–28] Our goal is to develop a decomposition of $1/r$ that is particularly suitable for use in MD simulations, and as a first step we establish some important properties of the BSA.

The absolute relative error of the BSA has the asymptotic bound

$$|r\mathcal{B}_b^\sigma(r) - 1| \lesssim 2^{3/2}\exp\left(-\frac{\pi^2}{2\ln(b)}\right) \quad (14)$$

as $b \to 1$ (see Appendix B for details). We denote this asymptotic error bound by $M_b$ and remark that it is independent of $\sigma$, being strictly a function of the base $b$. Numerical investigations show that $M_b$ (14) is within 1% of the exact bound for $b \leq 2$. The relative error for the BSA with $b = 2$ is illustrated in Fig. 1, with the asymptotic error bound $M_2$ plotted in red.

The BSA features the scaling property

$$\mathcal{B}_b^\sigma(r) = b\mathcal{B}_b^\sigma(br), \tag{15}$$

which follows immediately from the substitution $j = j' + 1$ in Eq. (11). It follows that the relative error has the property

$$r\mathcal{B}_b^\sigma(r) - 1 = (br)\mathcal{B}_b^\sigma(br) - 1. \tag{16}$$

Owing to Eq. (16), the relative error is uniquely determined as a function of $r$ by its values on the interval $[1,b)$. These values are repeated in the interval $[b^k, b^{k+1})$ for each integer $k$. The relative error is thus uniform across the entire real axis, being no worse than it is on the interval $[1,b)$. This uniform approximation property is shown in Fig. 1, which also illustrates that the BSA relative error has infinitely many roots that accumulate both to the origin and to infinity. The scaling invariance expressed by Eq. (16) implies this statement provided that there is at least one root in the interval $[1,b)$, which we prove in Appendix D.

**B. The BSA decomposition**

Ewald's method evaluates the near part in real space and the far part in $k$-space, where they are respectively localized. The Gaussian terms in the BSA decay super-exponentially (i.e., faster than exponentially) in real space and $k$-space as functions of $r / (b^j\sigma)$ and $k\sigma / b^j$, respectively (see Appendix C for the $k$-space form of the BSA). The BSA thus provides a natural approximate decomposition of $1 / r$ (the *BSA decomposition*) into terms with positive and negative $j$:

$$\mathcal{N}_{\text{BSA},b}^{\sigma}(r) \triangleq 2\ln(b) \sum_{j=-\infty}^{-1} b^{-j} G_\sigma(b^{-j} r) \tag{17}$$

$$\mathcal{F}_{\text{BSA},b}^{\sigma}(r) \triangleq 2\ln(b) \sum_{j=0}^{\infty} b^{-j} G_\sigma(b^{-j} r). \tag{18}$$

The components of the BSA decomposition are illustrated in Fig. 2.

The BSA decomposition has several desirable properties. The near and far terms have fast decay in real and $k$-space, respectively. The relative error of the BSA series converges exponentially in the limit $b \to 1$, as shown in Eq. (14). Finally, the Gaussian components of the BSA are separable; in fact, Gaussians are the only separable spherically symmetric functions. As will be discussed in Section V, this enables the use of lower-latency parallel algorithms if the infinite series can be truncated.

## C. The u-series

Two properties of the BSA decomposition make it not ideally suited for direct use in MD simulations. First, the near part of the BSA decomposition is not range limited, which leads to truncation errors when the near term is evaluated only up to a cutoff radius $r_c$. Second, the error of the BSA decomposition is large in absolute terms for values of $r$ much smaller than $r_c$, since the relative BSA error extends all the way down to $r = 0$. The $u$-series addresses these issues by retaining the far part of the BSA decomposition

$$\mathcal{F}_b^\sigma(r) \triangleq \mathcal{F}_{\text{BSA},b}^\sigma(r) = 2\ln(b) \sum_{j=0}^{\infty} b^{-j} G_\sigma(b^{-j} r) \tag{19}$$

while replacing the near part with

$$\mathcal{N}_b^\sigma(r) \triangleq \begin{cases} r^{-1} - \mathcal{F}_b^\sigma(r), & \text{if } r < r_c, \\ 0, & \text{if } r \geq r_c. \end{cases} \tag{20}$$

The cutoff radius $r_c$ is chosen to be the smallest root of

$$r\mathcal{F}_b^\sigma(r) - 1. \tag{21}$$

The $u$-series $\mathcal{S}_b^\sigma(r)$ is the sum of the near and far parts:

$$\mathcal{S}_b^\sigma(r) \triangleq \mathcal{N}_b^\sigma(r) + \mathcal{F}_b^\sigma(r). \tag{22}$$

By construction, the $u$-series equals $1/r$ exactly for $r < r_c$, and the near part has its range limited to $r < r_c$. For $r \geq r_c$, the relative error of the $u$-series is equal to that of $\mathcal{F}_b^\sigma(r)$, thus approaching the uniformly bounded relative error of the BSA $\mathcal{B}_b^\sigma(r)$ for $r \gg r_c$. Owing to the fast decay of $G_\sigma(r)$, the relative error of the BSA and the $u$-series approach each other rapidly when $r \geq r_c$, as illustrated in Fig. 3.

By choosing $r_c$ to be a root of $r\mathcal{F}_b^\sigma(r) - 1$, we ensure continuity of $\mathcal{S}_b^\sigma(r)$ at $r_c$, and thus along the entire real axis. As a result, the $u$-series does not incur the truncation errors associated with the cut-and-shift approach. The smallest root is chosen because it minimizes the $O(r_c^3)$ computational work required to compute the near term.

The definition of the $u$-series relies on the existence of roots of $r\mathcal{F}_b^\sigma(r) - 1$. We can confirm their existence by observing again that the near part of the BSA decomposition decays super-exponentially in $r$, so the relative error $r\mathcal{F}_b^\sigma(r) - 1$ of the far part approaches $r\mathcal{B}_b^\sigma(r) - 1$ for sufficiently large $r$, and thus has infinitely many roots away from the origin, as illustrated in Fig. 3. Because the near part is missing, the roots of $r\mathcal{F}_b^\sigma(r) - 1$ do not accumulate at the origin, as they do for $r\mathcal{B}_b^\sigma(r) - 1$. Fig. 3 plots the first four roots of $r\mathcal{F}_b^\sigma(r) - 1$.

Note that

$$\mathcal{F}_b^\sigma(r) = \frac{1}{\sigma}\mathcal{F}_b^1(r/\sigma), \qquad (23)$$

so that $r_c$ is equivalently the smallest root of $(r/\sigma)\mathcal{F}_b^1(r/\sigma) - 1$. For a given base $b$, define $r_0$ to be the smallest root of $r\mathcal{F}_b^1(r) - 1$; then we have the relation $r_c = r_0\sigma$. In practice, depending on the best approach for maximizing computational performance on a specific platform, one can either choose $\sigma$ and then set $r_c = r_0\sigma$, or choose $r_c$ and then set $\sigma = r_c / r_0$.

The smoothness of the $u$-series at $r_c$ can be increased by modifying the definition of the far part. We use the nomenclature $C^k$ *construction* for a family of $u$-series decompositions with $k$ continuous derivatives at all $r > 0$; accordingly the simplest form of the $u$-series—with far part defined in Eq. (19)—is referred to as the $C^0$ *construction*. A $C^1$-continuous form of the $u$-series that we refer to as the $C^1$ *construction*, and which numerical tests described in later sections suggest is particularly useful for MD applications, is obtained by varying the coefficient of the narrowest Gaussian in the far part of the series as follows:

$$\mathcal{F}_b^\sigma(r) = 2\ln(b)\left[w_0 G_\sigma(r) + \sum_{j=1}^\infty b^{-j} G_\sigma(b^{-j}r)\right]. \qquad (24)$$

To ensure that the resulting decomposition is $C^1$-continuous for a given base $b$, we determine the pair ($r_0$, $w_0$) for which $\mathcal{F}_b^1(r)$ and its first derivative evaluated at $r_0$ are equal to $1/r_0$ and $-1/r_0^2$, respectively. Using Eq. (24) we can solve for $w_0$ in terms of $r_0$ such that $\mathcal{F}_b^1(r_0) = 1/r_0$:

$$w_0(r_0) = \frac{1}{G_1(r_0)} \left[ \frac{1}{2\ln(b)r_0} - \sum_{j=1}^{\infty} b^{-j} G_1(b^{-j} r_0) \right]. \tag{25}$$

$r_0$ is then obtained by solving the equation

$$\frac{d}{dr} \mathcal{F}_b^1(r) \Big|_{r=r_0} = \frac{-1}{r_0^2}. \tag{26}$$

As is true for the $C^0$ construction, there exist multiple solutions to Eq. (26), and we pick $r_0$ to be the smallest one. The many solutions possible for any given $b$ give rise to the curves shown in Fig. 4. The $C^1$ construction contains some solutions that are not optimal; that is, there is another solution with both smaller $b$ (higher accuracy) and smaller $r_0$ (lower computational cost). The solid red curves in Fig. 4 highlight the set of all optimal $C^1$ constructions, which comprise a countably infinite set of intervals.

The points of discontinuity marked by bullets in Fig. 4 are points where two roots of Eq. (26) coalesce. For these special values of $b$, the $u$-series is $C^2$ continuous. The points of $C^2$ continuity can be found by solving the system of equations

$$\begin{cases} \mathcal{F}_b^1(r_0) = \frac{1}{r_0} \\ \frac{d}{dr} \mathcal{F}_b^1(r) \Big|_{r=r_0} = \frac{-1}{r_0^2}, \\ \frac{d^2}{dr^2} \mathcal{F}_b^1(r) \Big|_{r=r_0} = \frac{2}{r_0^3}, \end{cases} \tag{27}$$

with unknowns $b$, $r_0$, and $w_0$, where $\mathcal{F}_b^{\sigma=1}$ is the $C^1$ construction term specified by Eq. (24). Eq. (27) admits countably many solutions, forming a sequence such that $r_0 \to \infty$ as $b \to 1$. Each $C^2$ solution is the most accurate $u$-series in its interval. Parameters for the first three are given in Table 1, along with the corresponding bounds $M_b$ on the relative error.

Fig. 5 plots the $u$-series relative error bound $M_b$ against $r_0$, directly illustrating the tradeoff between computation (which increases with $r_0$) and error (which decreases with $r_0$). The gap in $r_0$ corresponds to the gap between the first two red curves in Fig. 4. Within the first interval, the $u$-series error decays roughly exponentially with $r_0$.

We believe that the $C^1$ construction suffices for all purposes relevant to MD simulations, especially since it has a subset of $C^2$-continuous solutions that can be used if such solutions are required. For even smoother approximations, additional parameters could be varied, for example the width $s_0$ of the narrowest Gaussian, as in

$$\mathcal{F}_b^\sigma(r) = 2\ln(b)\left[w_0 G_{\sigma s_0}(r) + \sum_{j=1}^\infty b^{-j} G_\sigma(b^{-j} r)\right]. \tag{28}$$

Varying the parameters defining the narrowest Gaussian is preferable, in order to prevent the appearance of large errors away from $r_0$. We refer to Eq. (28) as the *$C^2$ construction* and provide parameters for $b = 2$ in the third row of Table 2, but we found that this and smoother constructions suffer from larger values of $r_0$, as well as larger overshoot or undershoot of the BSA relative error. With respect to the magnitude of the relative error for $r > r_0$, the $C^1$ construction is well behaved, as illustrated in Fig. 6 for $b = 2$. Although it undershoots the lower bound $-M_2$ near $r_0$, it blends more smoothly at $r_0$ than the $C^0$ construction (compare to the black solid line of Fig. 3, which does not go out of bounds, but exhibits a clear kink at $r_0$). For $b \le 2$, numerical investigations show that the overshoot/undershoot never exceeds 20%, in part because the optimized values for the parameter $w_0$ remain close to 1 (see Fig. 7).

## IV. Numerical Experiments

We now assess the accuracy and performance of the *u*-series. In Section IVA we calculate errors in the energy and pressure obtained using *u*-series $C^1$ and $C^2$ constructions for 100 configurations obtained from the simulation of a lipid bilayer. We compare these errors to those obtained for the same set of configurations using an Ewald decomposition. This allows us to establish parameters that give rise to similar accuracy between the two decomposition methods, and hence to estimate their relative performance. Then, in Section IVB, we perform a series of simulations using different *u*-series parameters, and calculate observables that are known to be sensitive to the electrostatics. These results help us recommend *u*-series parameters that are useful in practice for MD simulations.

The lipid bilayer system we simulated was composed of 72 dipalmitoylphosphatidylcholine (DPPC) lipids solvated by 2189 molecules of water. The simulations were performed at a temperature of 323 K, a pressure of 1 atm, and a surface tension of 0 dynes cm$^{-1}$. Under these conditions, the bilayer is in its fluid state,[29] and the aspect of the box fluctuates by a few percentage points. These large fluctuations couple to the highly non-homogeneous distribution of charges in the direction perpendicular to the bilayer, making DPPC a sensitive test for electrostatics. Indeed, simple truncation of Coulomb interactions is known to produce major artifacts in DPPC,[30] and, although smoothly tapering $1/r$ to zero before truncation can alleviate some of the problems, there can still be large discrepancies in properties such as the electrostatic potential difference between the center of the bilayer and the surrounding water.[31] The 100 simulation configurations used in section IVA were sampled at 1-ns intervals from a 100-ns simulation that used a very accurate parameterization of the Ewald decomposition. The simulations described in section IVB used the *u*-series. In all our simulations, the cutoff radius for van der Waals interactions was 12 Å.

## A. Comparison to the Ewald decomposition

We start by focusing on errors that are intrinsic to the decomposition scheme: those that remain even if the direct and *k*-space sums are performed exactly.

In Fig. 8 we show the absolute error when $\sigma = 1$ and $r_c \approx 1.989$ for the Ewald decomposition and the $C^1$-continuous *u*-series with base $b = 2$. The absolute error Ewald($r$) – 1 / $r$ is constant and large in the region $r \leq r_c$. Using a switching function to taper the near Ewald potential at $r_c$ does not decrease the maximum interaction error. Beyond the cutoff, the error decays as fast as a Gaussian of width $\sigma$. The absolute error $S_2^1(r)$ – 1 / $r$ for the *u*-series decays less rapidly, but is much smaller in magnitude than the Ewald error for small values of *r*.

To see how these errors in approximating 1 / *r* translate into errors in the computation of energies and pressures for configurations obtained from MD simulations, we used 100 configurations from a simulation of the bilayer system described above. We computed the energy and pressure for these configurations using different decomposition schemes and parameterizations, including a very accurate Ewald parameterization that yields essentially exact results. The absolute errors in the energies and pressures for the various parameterizations were then calculated as differences from these essentially exact results. In order to isolate errors due to the decomposition, a very accurate grid-based method with a very fine grid (128 × 128 × 128) was used to evaluate the far part, and all computations were carried out in double precision.

The results of our computations, which all used $r_c = 8$ Å, are shown in Fig. 9. The red lines show the energy and pressure errors for the $C^1$-continuous $u$-series for the same values of $b = 2$ and $r_0 = 1.989$ used in Fig. 8, so that $\sigma = 8 / 1.989 \approx 4.02$ Å. Such a value of $\sigma$ leads to large absolute errors for the Ewald decomposition (not shown), but these errors can be decreased by decreasing $\sigma$. We find that we must decrease $\sigma$ to approximately 2.50 Å for the Ewald decomposition errors to match the magnitude of the errors characteristic of the $C^1$-continuous $u$-series with $b = 2$ (see Fig. 9). The value of $\sigma$ for the $u$-series is $4.02 / 2.50 \approx 1.61$ times larger, which results in substantial computational savings, as we discuss below. If greater accuracy is required, then a $u$-series with a smaller base $b$ is a good choice: The $C^2$-continuous $u$-series with $b \approx 1.63$ and $\sigma = 2.91$ Å shown in Fig. 9, for example, has an error that is one order of magnitude smaller than both Ewald and $u$-series with $b = 2$, $\sigma = 4.02$ Å, yet is still more computationally efficient than Ewald ($\sigma = 2.91$ Å $> 2.50$ Å).

So far we have focused on the errors intrinsic to the decomposition scheme by computing the far part very accurately. In practice, however, such accurate computations would be prohibitively expensive, so faster but more approximate computations would be performed leading to another source of error. We now redo the calculations leading to Fig. 9, but evaluate the far part using the level of accuracy that we typically use in MD simulations, and show the results in Fig. 10. Specifically, we use the GSE[15] method, which was originally described in the context of the Ewald decomposition, but which is also straightforward to use to calculate the far contribution of the $u$-series decomposition. In GSE, charges are spread from particles to grid points within a certain radius; the tradeoff between speed and accuracy is principally controlled by the ratio of this charge-spreading cutoff radius to the grid spacing. We have found through experience that a reasonable compromise between speed and accuracy is obtained when this ratio is 3.8, a value we have used extensively in simulations of a variety of systems, and which we also choose here. Other GSE parameters follow from this choice: The width of the spreading Gaussian, which doesn't affect speed and so can be chosen to optimize accuracy, is $\sigma_s = 0.83\Delta$, where $\Delta$ is the grid spacing; the GSE requirement $\sigma \geq \sqrt{2}\sigma_s$ yields a maximum allowable grid spacing of $\Delta \approx \sigma / 1.17$.

For our DPPC system, this ratio leads to a 24 × 24 × 32 grid for the Ewald decomposition, and 15 × 15 × 21 and 21 × 21 × 28 grids for the $u$-series decomposition with $b = 2$ and $b \approx 1.63$, respectively. Fig. 10 plots the energy and pressure errors using these more practical computations of the far contributions. The results are nearly identical to those in Fig. 9, validating our finding that the $u$-series is more computationally efficient than Ewald for this system. Not surprisingly, the most accurate method in the figures ($u$-series with $b \approx 1.63$) suffers most from the error related to the coarser grids: The biggest difference between Figs. 9 and 10 is a small but noticeable increase in energy error for this simulation.

Based on the results of this section and the arguments about computational work scaling in Section II, we now estimate the ratio of total work required to compute the electrostatic interaction with comparable error using $u$-series and Ewald decompositions under certain assumptions. Because they decay super-exponentially with $r_0 = r_c / \sigma$ and only polynomially with $r_c$ for both Ewald and $u$-series, the errors are primarily functions of $r_0$. In Section II, we argued that the minimum number of operations required to evaluate the direct and $k$-space sums are proportional to at least $r_c^3$ and $1 / \sigma^3$, respectively; the total number of operations required is thus

$$N_{\text{ops}} \cong C_n r_c^3 + C_f r_0^3 / r_c^3, \tag{29}$$

for some positive constants $C_n$ and $C_f$. Choosing $r_c$ to minimize $N_{\text{ops}}$ yields

$$\min_{r_c} N_{\text{ops}} \cong 2\sqrt{C_n C_f} r_0^{3/2}. \tag{30}$$

If polynomial interpolation is used to compute the interaction functions, then the overall computational cost is largely independent of the specific near/far functional forms (Eq. (6) for Ewald vs. Eqs. (19) and (20) for *u*-series). Both Ewald and the *u*-series thus obey Eq. (30) with roughly the same values of $C_n$ and $C_f$, and differ only in the value of $r_0$. Under this assumption, the observed ratio $r_{0,\text{ewald}} / r_{0,u\text{-series}} = 1.61$ for $r_c = 8$ implies that the *u*-series requires about $1.61^{-3/2} \approx 1/2$ of the computational effort for equivalent accuracy. We emphasize that this estimate of the ratio of *u*-series to Ewald computational work makes a number of simplifying assumptions, including perfect balancing of the computational load between the near and far terms, and computational work that is strictly proportional to $r_c^3$ and $1/\sigma^3$, with equal constant factors. In practice, the relative performance of the two decompositions is dependent on the particular algorithms and computer architecture used.

### B. Choosing *u*-series parameters for use in simulations

Although errors in the energy and pressure of DPPC are a useful way to judge the fidelity of electrostatics calculations, it is not clear, a priori, how these errors will impact the quality of the simulations themselves. To assess the accuracy required to avoid substantial simulation artifacts, we looked at two properties mentioned earlier as being sensitive to electrostatics: the surface area of the bilayer and the electrostatic potential difference between the water and the center of the membrane. We used as our baseline an Ewald simulation with a large cutoff radius ($r_c = 13.0$ Å) to ensure high accuracy. As a threshold for acceptable accuracy, we required our observed mean properties to fall within the natural fluctuations—chosen here to be plus or minus one standard deviation—as estimated from the baseline run; this should help ensure that simulations using the electrostatic approximations still significantly overlap the target ensemble. Simulations were run for 2 to 4 µs each, with frames saved every 0.24 ns, and the first 120 ns of each simulation discarded from the analysis. Surface area was defined as the area of the simulation box parallel to the membrane (the $x$-$y$ plane), and the electrostatic potential drop was estimated by first using the tool pmepot[32] to compute a smoothed estimate of the electrostatic potential on a grid, and then averaging grid values that share the same $z$ coordinate.[31] All simulations used a 2.5 fs time step and a $64 \times 64 \times 64$ grid to compute the far electrostatics.

Fig. 11 shows the error in mean membrane properties as a function of $r_c$ for the $C^1$-continuous $u$-series with $b = 2$, $r_c / \sigma \approx 1.989$, and the $C^2$-continuous $u$-series with $b \approx 1.63$, $r_c / \sigma \approx 2.752$ (refer to Tables 1 and 2 for more precise parameter values). DPPC simulations exhibited natural fluctuations in surface area with a standard deviation of about 3%, and all of our simulations yielded errors less than half this value. Similarly, the mean potential drop (averaged across 40-ns windows) had a standard deviation of about 2.4%, again larger than the errors attributable to our choice of $r_c$, suggesting that our accuracy was sufficient. By comparison, simulations that use cutoff electrostatics can have errors in the mean potential drop of well over 100%,[31] and are likely too inaccurate for membrane systems. Since DPPC is the most sensitive realistic system we have tested with respect to electrostatic approximations, it seems reasonable to expect that even the less accurate $u$-series settings shown in Fig. 11 should be conservative enough for most simulation purposes. In fact, we have used fairly aggressive parameters ($b = 2$, $r_c / \sigma \approx 1.989$, $r_c = 8$ Å) in other tests—reversible folding of villin headpiece and the temperature-dependent helicity of an Ala$_{15}$ peptide, for example—and have achieved results consistent with simulations that use more accurate electrostatics. We also discuss parameter choice in the Summary and Conclusion section.

## V.   A Separable Real-Space Algorithm for Evaluating the Electrostatic Energy

In arriving at our main finding that the *u*-series requires roughly half the computation of the Ewald approach, we assumed that the far part of each decomposition would be calculated using the same algorithm. Since the *u*-series far part is well approximated by a sum of a small number of Gaussians, however, it can be computed by certain algorithms that the Ewald far part cannot, leading to an additional potential performance advantage for the *u*-series in certain situations. In this section, we sketch one such algorithm that may be advantageous on massively parallel machines.

On massively parallel supercomputers, the time required to evaluate the Ewald *k*-space sum is strongly affected by communication latency.[6,7] Since the far interaction is not range limited, any parallel algorithm requires at least one global operation with the property that data from each processor affects the result of the computation on every other processor. In practice, computing the far interaction with a single global operation is rarely achieved. Many MD packages use PME[11,12] to compute the *k*-space sum, in which a fast Fourier transform (FFT) is performed on the grid, followed by a multiplication in *k*-space, followed by an inverse Fourier transform. Each Fourier transform is a global operation, so PME requires two global operations.

The *u*-series *k*-space sum can alternatively be computed in real space with a single global operation by taking advantage of the dimensional separability of the Gaussian kernel:

$$e^{-\frac{r^2}{2\sigma^2}} = e^{-\frac{x^2}{2\sigma^2}} e^{-\frac{y^2}{2\sigma^2}} e^{-\frac{z^2}{2\sigma^2}}. \tag{31}$$

First, we truncate the far part of the *u*-series at $N$ terms, with $N$ chosen to be large enough so that the truncation error is acceptably small (this will be quantified shortly). As long as a separable function $S(\mathbf{r}) = S_0(x)S_1(y)S_2(z)$ such as a B-spline (PME) or Gaussian (GSE) is used to spread particle charges onto the grid, the *k*-space kernel for the resultant on-grid convolution

$$\frac{4\pi \ln(b)\sigma^2}{|\hat{S}(k)|^2} \left[ w_0 e^{-(\sigma^2)k^2/2} + \sum_{j=1}^{N-1} b^{2j} e^{-(b^{2j}\sigma^2)k^2/2} \right] \tag{32}$$

is a sum of separable functions, where $\hat{S}$ denotes the Fourier transform of $S$. In real space, the on-grid potential is the circular convolution of the on-grid charge with the inverse Fourier transform of this kernel. Alternatively, we may regard the on-grid potential as the sum of $N$ potentials, each obtained by convolving the on-grid charge with the inverse Fourier transforms of one of the Gaussians appearing in Eq. (32). The $N$ Gaussians are separable by dimension, as are their inverse Fourier transforms. The potential associated with a given Gaussian thus represents a separable three-dimensional convolution (i.e., the composition of three one-dimensional convolutions).

Each separable convolution is performed in three sequential rounds: In the first round, for every $(y, z)$ grid coordinate pair, grid values are broadcast along the $x$ dimension, and the $x$ convolution is computed. Each output point of the convolution can be computed independently with no additional communication, since the convolution is executed directly in real space. The same procedure is then repeated in the $y$ and $z$ dimensions. The three rounds of convolutions are performed for each of the $N$ *u*-series terms in parallel, and the results are summed at the end to produce the final grid potentials. In total, three sequential rounds of communication are required, as opposed to the six sequential rounds necessary to perform both forward and backward Fourier transforms.

Computing the *k*-space sum in real space as described requires greater total computation and bandwidth because separate sets of convolutions must be performed for each of the $N$ terms, whereas FFT-based approaches operate only on the sum of the terms. This tradeoff of increased computation and bandwidth for reduced latency may be advantageous on massively parallel machines. Additionally, the computation and bandwidth can be reduced using other techniques if necessary. The on-grid Gaussian kernels corresponding to different terms have increasing widths and, consequently, are more suitably handled on increasingly coarser grids. The *u*-series can thus be implemented at reduced computational cost by computing its far part using the multilevel summation method of Skeel et al.[33–35]

The practicality of the reduced-latency real-space algorithm depends critically on the value of $N$, because both bandwidth and computation scale in proportion to $N$. Fortunately, periodicity causes the Gaussians that are broader than the longest side of the unit cell to self-cancel, and we can show (Appendix F) that the relative error for the far part of the periodic energy when truncated to $N$ terms is bounded by

$$\exp\left(2N\ln(b) - 2(b^{2N} - 1)\frac{\pi^2 \sigma^2}{L^2}\right), \tag{33}$$

where $L = \max\{L_x, L_y, L_z\}$, generally leading to small values of $N$. Consider, as an example, the DPPC system, for which $L = 70$ Å. For the $C^1$-continuous *u*-series with $b = 2$, we have $\sigma \approx 4.02$ Å. The upper bound provided by Eq. (33) is $1.5 \times 10^{-5}$ for $N = 4$, and a tiny $1.1 \times 10^{-26}$ for $N = 5$. Given the typical accuracy required for MD simulations, $N = 4$ suffices for this particular system.

## VI. Summary and Conclusion

The *u*-series decomposes $1/r$ into a sum of near and far parts such that the resulting approximation has a number of desirable properties: It is smooth on the entire real axis, is exact up to a cutoff radius $r_c$, and has uniformly bounded relative error for $r > r_c$. These factors contribute to the *u*-series being more accurate than the Ewald decomposition for a given amount of computational effort. This in turn lends the *u*-series a computational advantage: In our experiments, roughly half as much work was required to achieve the same degree of accuracy as the Ewald decomposition.

The *u*-series is also constructed so that its far part is a sum of separable functions, and we have shown that, with periodic boundary conditions, only a small number of summation terms need to be evaluated in order to compute the far part to machine precision. This enables the use of minimal-latency algorithms wherein the three-dimensional convolutions with each of these Gaussians are computed in parallel in real space as a product of three one-dimensional convolutions. This algorithmic advantage of the *u*-series for massively parallel supercomputers will likely become more important over time as computational density and communication bandwidth continue to scale more rapidly than communication latency. Section V describes only one possible algorithm for exploiting the separability of the *u*-series (one that has been implemented in Anton 2),[36] leaving room for further algorithmic research.

Among the *u*-series variants that we have presented, the $C^1$ construction appears to be the most useful in practice for MD simulations of biophysical systems. Our experience has been that most simulations are well served by the $C^1$ construction with $b = 2$, which is sufficiently accurate even for electrostatic cutoff radii as low as $r_c = 8$ Å. The parameters for the $C^1$ construction with $b = 2$ are given by the second row of Table 2, and represent our recommended *u*-series parameterization for most MD simulations carried out near or above room temperature. When the accuracy of the electrostatics is the primary concern, the $C^2$ construction may be used; even the least accurate $C^2$ parameterization given in Table 1 ($b = 1.629…$) is more accurate than the Ewald decomposition with the same cutoff radius and has a comparable computational cost.

In this work, we have focused on the decomposition of the Coulomb potential $1/r$, but the bilateral series approximations in Section III can be used more generally for $1/r^\alpha$, as noted by Beylkin and Monzon.[21,22] This in turn leads to $u$-series decompositions for $1/r^\alpha$, which may prove useful, for example, for splitting the Lennard-Jones dispersion interaction $1/r^6$. Similarly, although the present work focuses exclusively on decompositions based on Gaussian series approximations, Appendix A shows how the BSA can be constructed based on series of arbitrary well-behaved functions $\phi$ not limited to Gaussians, leading to alternative formulations. Although Gaussians have several desirable properties for our purposes, formulations based on alternative functions may be useful in other contexts.

The present work leaves open some mathematical questions about the $u$-series that would benefit from further investigation. We have shown that the pointwise convergence of the $u$-series is exponentially fast as $b$ approaches 1 at constant $\sigma$. We have not proven, however, that this convergence for the Coulomb kernel implies similarly fast convergence for the total energy. While pointwise convergence typically entails convergence of integrals, a more precise statement and formal proof of the convergence of the total energy would be desirable. Additionally, for constant $b$, the $u$-series converges to $1/r$ in the limit $r_c \to \infty$. The numerical evidence presented in Appendix E suggests that the total electrostatic energy converges polynomially fast, with a degree one larger than the degree of smoothness of the $u$-series, but we do not have a proof for this conjecture. Mathematical investigations beyond the scope of this paper are thus warranted.

## Acknowledgments

The authors thank Michael Eastwood and Cliff Young for helpful early discussions, and Mollie Kirk and Berkman Frank for editorial assistance.

This study was conducted and funded internally by D. E. Shaw Research, of which D.E.S. is the sole beneficial owner and Chief Scientist and with which all the authors are affiliated.

## Appendix A.  Convergence of the BSA to $1/r^\alpha$

Let $\alpha > 0$ be a positive number and $\phi(r)$ a continuous, piecewise-smooth function (the *scaling function*) decaying sufficiently rapidly at infinity.  Let

$$\tilde{\phi}(\alpha) = \int_0^\infty \phi(u) u^{\alpha-1} du \qquad (A1)$$

be the Mellin transform of $\phi$.  For any base $b > 1$, the BSA is defined by

$$\mathcal{B}_{\alpha,b}(r) = \frac{\ln(b)}{\tilde{\phi}(\alpha)} \sum_{j=-\infty}^{\infty} b^{-\alpha j} \phi(b^{-j} r). \qquad (A2)$$

The BSA converges to $1/r^\alpha$ in the limit $b \to 1$, and this can be established as follows.  Let $f(x) = b^{-\alpha x} \phi(b^{-x} r)$.  The Poisson summation formula states that the BSA series is equal to

$$\frac{\ln(b)}{\tilde{\phi}(\alpha)} \sum_{j=-\infty}^{\infty} f(j) = \frac{\ln(b)}{\tilde{\phi}(\alpha)} \sum_{n=-\infty}^{\infty} \int_{-\infty}^{\infty} f(x) e^{-i2\pi n x} dx.$$

Perform the substitution $u = b^{-x} r$ and set $b = \exp(\Delta)$ to express the BSA as

$$\frac{1}{\tilde{\phi}(\alpha)r^\alpha}\sum_{n=-\infty}^{\infty}\int_0^\infty u^{\alpha-1}\phi(u)u^{i2\pi n/\Delta}e^{-i2\pi n\ln(r)/\Delta}du = \frac{1}{r^\alpha}+\frac{1}{r^\alpha}\sum_{n\neq 0}\frac{\tilde{\phi}(\alpha+i2\pi n/\Delta)}{\tilde{\phi}(\alpha)}e^{-i2\pi n1\ (r)/\Delta}.$$

This equation gives an explicit formula for the relative error, namely

$$r^\alpha \mathcal{B}_{\alpha,b}(r) - 1 = \sum_{n\neq 0}\frac{\tilde{\phi}(\alpha+i2\pi n/\Delta)}{\tilde{\phi}(\alpha)}e^{-i2\pi n1\ (r)/\Delta}, \tag{A3}$$

which is bounded by

$$\left|r^\alpha \mathcal{B}_{\alpha,b}(r) - 1\right| \leq \sum_{n\neq 0}\frac{|\tilde{\phi}(\alpha+i2\pi n/\Delta)|}{\tilde{\phi}(\alpha)}. \tag{A4}$$

The Mellin transform of a function with $k$ derivatives of bounded variation decays as fast as $O(x^{-k-1})$ in the complex plane along the imaginary direction (a precise statement can be inferred from Theorem 2.4 (a) of ref. 37). If $\phi(r)$ is piecewise smooth and its derivative has bounded variation, then the r.h.s. sum in Eq. (33) is convergent and vanishes in the limit $\Delta \to 0$ (i.e., $b \to 1$). Consequently, $r^\alpha \mathcal{B}_{\alpha,b}(r) \to 1$ as $b \to 1$, and the convergence claim is validated.

## Appendix B. BSA Error Bounds

When $\phi$ is a Gaussian of width $\sigma$, Eq. (A4) can be written explicitly as

$$\left|r^\alpha \mathcal{B}^\sigma_{\alpha,b}(r) - 1\right| \leq \sum_{n\neq 0}\frac{|\Gamma(\alpha/2+i\pi n/\Delta)|}{\Gamma(\alpha/2)}.$$

As $\Delta \equiv \ln(b) \to 0$ or as $n$ increases, $|\Gamma(\alpha/2 + i\pi n/\Delta)|$ decays exponentially. The r.h.s. sum above becomes dominated by the $n = -1$ and $n = 1$ terms, which have equal magnitudes. The asymptotic decay can be established with the help of Stirling's formula, which asserts

$$\Gamma\left(\tfrac{\alpha}{2}+i\tfrac{\pi}{\Delta}\right) \sim \sqrt{2\pi}\left(\tfrac{\alpha}{2}+i\tfrac{\pi}{\Delta}\right)^{\frac{\alpha-1}{2}+i\frac{\pi}{\Delta}} e^{-\alpha/2-i\pi/\Delta},$$

as $\Delta \to 0$. The absolute value of the r.h.s. is

$$\sqrt{2\pi}\left(\tfrac{\alpha^2}{4}+\tfrac{\pi^2}{\Delta^2}\right)^{\frac{\alpha-1}{4}} \exp\left(-\tfrac{\alpha}{2}-\tfrac{\pi}{\Delta}\arctan\left(\tfrac{2\pi}{\Delta\alpha}\right)\right). \tag{B1}$$

For positive values of $x$, we have the identity

$$\arctan(x) = \tfrac{\pi}{2} - \arctan\left(\tfrac{1}{x}\right) = \tfrac{\pi}{2} - \tfrac{1}{x} + O\left(\tfrac{1}{x^2}\right),$$

which implies

$$-\tfrac{\alpha}{2} - \tfrac{\pi}{\Delta}\arctan\left(\tfrac{2\pi}{\Delta\alpha}\right) = -\tfrac{\pi^2}{2\Delta} + O(\Delta).$$

Retaining only the dominant terms in Eq. (B1), we infer

$$\left|\Gamma\left(\tfrac{\alpha}{2}+i\tfrac{\pi}{\Delta}\right)\right| \sim (2\pi)^{1/2}\left(\tfrac{\alpha^2}{4}+\tfrac{\pi^2}{\Delta^2}\right)^{\frac{\alpha-1}{4}} \exp\left(-\tfrac{\pi^2}{2\Delta}\right), \tag{B2}$$

in the limit $\Delta \to 0$. Adding the equal contribution of $|\Gamma(\alpha/2 - i\pi/\Delta)|$ and dividing by $\Gamma(\alpha/2)$ produces the asymptotic bound

$$\left|r^\alpha \mathcal{B}^\sigma_{\alpha,b}(r) - 1\right| \lesssim \tfrac{2(2\pi)^{1/2}}{\Gamma(\alpha/2)}\left(\tfrac{\alpha^2}{4}+\tfrac{\pi^2}{\Delta^2}\right)^{\frac{\alpha-1}{4}} \exp\left(-\tfrac{\pi^2}{2\Delta}\right). \tag{B3}$$

This generalizes the bound given for $\alpha = 1$ in Eq. (14).

Numerical investigations for $b \leq 2$ have shown that this bound is a reliable estimate of the relative error maximum, meaning that it is either an upper bound (for $\alpha > 3/2$), or within 1% of the exact result (for $\alpha \leq 3/2$). The excellent agreement for small values of $\alpha$ is illustrated in Fig. 1 for Coulomb potentials. The maximum relative error is about $2 \times 10^{-3}$ for $b = 2$ and less than $2 \times 10^{-6}$ for $b = 2^{1/2}$.

## Appendix C. The BSA in Fourier Space

For the Coulomb kernel considered in the body of the paper, the expression for the BSA in $k$-space is given by

$$\widehat{\mathcal{B}}_b^\sigma(\mathbf{k}) \equiv \frac{1}{(2\pi)^{3/2}} \int_{\mathbb{R}^3} \mathcal{B}_b^\sigma(r) e^{-i\mathbf{k}\cdot\mathbf{r}} d\mathbf{r} = \frac{2\ln(b)\sigma}{\sqrt{2\pi}} \sum_{j=-\infty}^{\infty} \frac{1}{b^{2j}} \exp\left[-\frac{1}{2}\left(\frac{k\sigma}{b^j}\right)^2\right], \tag{C1}$$

where $\widehat{\phantom{x}}$ denotes Fourier transform. Eq. (C1) closely resembles the BSA itself in Eq. (11); we can generalize this result to arbitrary scaling functions $\phi(r)$ and potentials of the form $1/r^\alpha$.

Observe first that the three-dimensional Fourier transform

$$\hat{\phi}(\mathbf{k}) = \frac{1}{(2\pi)^{3/2}} \int_{\mathbb{R}^3} \phi(r) e^{-i\mathbf{k}\cdot\mathbf{r}} d\mathbf{r} = \sqrt{\frac{2}{\pi}} \frac{1}{|\mathbf{k}|} \int_0^\infty r\phi(r) \sin(|\mathbf{k}|r) dr \tag{C2}$$

of the scaling function $\phi(r)$ is spherically symmetric in $k$-space. As a reminder of this symmetry, we shall denote the Fourier transform by $\hat{\phi}(k)$, even though the proper notation is $\hat{\phi}(\mathbf{k})$. The Fourier transform of the r.h.s. of Eq. (C2) equals

$$\frac{\ln(b)}{\tilde{\phi}(\alpha)} \sum_{j=-\infty}^{\infty} b^{-(\alpha-3)j} \hat{\phi}(b^j k) \stackrel{j \mapsto -j}{=} \frac{\tilde{\hat{\phi}}(3-\alpha)}{\tilde{\phi}(\alpha)} \left[\frac{\ln(b)}{\tilde{\hat{\phi}}(3-\alpha)} \sum_{j=-\infty}^{\infty} b^{-(3-\alpha)j} \hat{\phi}(b^{-j}k)\right]. \tag{C3}$$

Clearly, the expression in the square brackets is a BSA in $k$ converging to $1/k^{3-\alpha}$ as $k \to \infty$, with an error bound that decays exponentially in $b$. When $\phi$ is a Gaussian of width $\sigma$, $\hat{\phi}$ is a Gaussian of width $1/\sigma$, and setting $\alpha = 1$ leads to Eq. (C1). More generally, we show that

$$f(\alpha) \triangleq \frac{\tilde{\hat{\phi}}(3-\alpha)}{\tilde{\phi}(\alpha)} = \frac{\int_0^\infty \hat{\phi}(k) k^{(3-\alpha)-1} dk}{\int_0^\infty \phi(u) u^{\alpha-1} du} = \frac{(2\pi)^{3/2}}{c_\alpha}, \tag{C4}$$

where

$$c_\alpha \triangleq \frac{2^\alpha \pi^{3/2} \Gamma(\alpha/2)}{\Gamma((3-\alpha)/2)}.$$

For this to hold, it suffices to make the following assumptions about $\phi(r)$. First, $\phi(r)$ decays at infinity at least as fast as $r^{-3}$, together with its first two derivatives. Second, $\phi(r)$ has second-order derivatives of bounded variation. Note that a Gaussian satisfies these assumptions. As can be demonstrated by integration by parts of Eq. (C2), the conditions stated are such that the three-dimensional Fourier transform $\hat{\phi}(k)$ decays at least as fast as $k^{-3}$.

With these assumptions, the integrals appearing in Eq. (C4) are properly defined for all $\alpha$ in the complex strip $0 < \Re(\alpha) < 3$. Being Mellin transforms, the integrals are analytic on the mentioned strip. As a ratio of holomorphic functions with the denominator not identically zero, $f(\alpha)$ is meromorphic on the complex strip $0 < \Re(\alpha) < 3$. For $2 < \Re(\alpha) < 3$, by Eq. (C2) and absolute integrability, we compute

$$\int_0^\infty \hat{\phi}(k) k^{2-\alpha} dk = \sqrt{\frac{2}{\pi}} \int_0^\infty \int_0^\infty k^{1-\alpha} \phi(r) r \sin(kr) dr dk \stackrel{k=x/r}{=} \sqrt{\frac{2}{\pi}} \left[\int_0^\infty \phi(r) r^{\alpha-1} dr\right]\left[\int_0^\infty x^{1-\alpha} \sin(x) dx\right].$$

It follows that

$$f(\alpha) = \sqrt{\frac{2}{\pi}} \int_0^\infty x^{1-\alpha} \sin(x)\, dx.$$

By Eq. (C2), the last integral is the three-dimensional Fourier transform of $r^{-\alpha}$ evaluated at any point $\mathbf{k}$ with $|\mathbf{k}| = 1$. It is well known that the three-dimensional Fourier transform of $r^{-\alpha}$ is

$$\frac{(2\pi)^{3/2}}{c_\alpha k^{3-\alpha}},$$

whence $f(\alpha) = (2\pi)^{3/2} / c_\alpha$. The equality $f(\alpha) = (2\pi)^{3/2} / c_\alpha$ is extended to the entire strip $0 < \Re(\alpha) < 3$ by analytic continuation, and the proof of Eq. (C4) is concluded.

## Appendix D. Existence of a Root in Eq. (21)

To prove the existence of a root of the BSA relative error $r^\alpha \mathcal{B}_{\alpha,b}(r) - 1$ in the interval $[1, b)$, consider its integral against the scale-invariant measure $dr / r$,

$$\int_1^b \left(r^\alpha \mathcal{B}_{\alpha,b}(r) - 1\right) r^{-1} dr = \int_0^\Delta \left(e^{\alpha x} \mathcal{B}_{\alpha,b}(e^x) - 1\right) dx, \tag{D1}$$

where $b = \exp(\Delta)$. Substituting the r.h.s. of Eq. (33) in Eq. (D1) produces

$$\int_0^\Delta \sum_{n \neq 0} \frac{\tilde{\phi}(\alpha + i2\pi n/\Delta)}{\tilde{\phi}(\alpha)} e^{-i2\pi n x / \Delta}\, dx. \tag{D2}$$

We showed in Appendix A that the sum in Eq. (D2) is absolutely convergent, so by Fubini's theorem we may interchange the sum and integral,

$$\sum_{n \neq 0} \frac{\tilde{\phi}(\alpha + i2\pi n/\Delta)}{\tilde{\phi}(\alpha)} \int_0^\Delta e^{-i2\pi n x / \Delta}\, dx = 0. \tag{D3}$$

The integral of the relative error against the positive measure $dr/r$ could not vanish if the relative error did not change sign on the interval $[1, b)$. There thus exists at least one root in the interval $[1, b)$, and the root structure illustrated in Fig. 1 is established as a universal property of the bilateral series.

## Appendix E. Impact of $r_c$ on $u$-series vs. Ewald

In Section IV we saw numerical evidence that reducing $\sigma$ by a factor of 1.61 (Fig. 8) produces an Ewald decomposition similar in accuracy to the $C^1$-continuous $u$-series with $b = 2$ and $r_c = 8$ Å. For larger values of $r_c$, the factor 1.61 needs to be increased. Fig. 12 plots the root-mean-square deviations (RMSDs) of the errors as a function of the cutoff $r_c$, subject to the same ratios $r_c/\sigma$ used for Fig. 9. The errors are calculated for the same 100 lipid bilayer configurations used in Section IV. As $r_c$ increases, the energy fluctuations for the different methods decay to 0 at rates that are related to their smoothness. Fig. 12 shows that the RMSDs for the errors in the electrostatic energy decay as fast as $1/r_c$ for the Ewald decomposition with constant $r_c/\sigma$ (which is $C^0$ continuous), $1/r_c^2$ for the $C^1$-continuous $u$-series with $b = 2$, and $1/r_c^3$ for the $C^2$-continuous $u$-series with $b \approx 1.63$. This figure supports the general postulate that the errors for a $C^d$-continuous $u$-series decrease as fast as $1/r_c^{d+1}$.

Based on the results of the preceding paragraph, it is tempting to consider smoother $u$-series than those provided by the $C^1$ construction. For $b = 2$, the $C^2$-continuous $u$-series obtained from Eq. (28) has $r_0 \approx 2.394$, a considerable increase from the corresponding values of 1.989 and 1.843 for the $C^1$- and $C^0$-continuous $u$-series of same base $b$ (see Table 1). In Fig. 13, we compare this $C^2$-continuous $u$-series to the $C^1$-continuous $u$-series having the same $r_0 = r_c/\sigma$ (that is, the same computational effort) rather than the same base $b$. The $C^1$-continuous $u$-series appears to be about as good as the $C^2$-continuous *u*-series for moderate values of $r_c$ and better for smaller values. It will eventually be overtaken by the $C^2$-continuous *u*-series in the limit of large $r_c$, but the values of $r_c$ for which this happens are so large that they are impractical. We believe this result suggests that the $C^1$ construction is sufficiently versatile to accommodate all needs arising in the practice of MD simulations, and we choose not to explore smoother approximations any further.

## Appendix F. Error Bound on Distant Energy in Terms of Series Length

For the *u*-series decomposition, the far part of the periodic electrostatic energy (before subtracting the self-interactions) is

$$2\pi \ln(b) \sigma^2 V \sum_{\mathbf{k} \neq 0} |C_\mathbf{k}|^2 \sum_{j=0}^{N-1} b^{2j} e^{-b^{2j}|\mathbf{k}|^2 \sigma^2/2}, \tag{F1}$$

where $\mathbf{k}$ are the reciprocal lattice vectors from Eq. (4). The terms with $j \geq N$ in Eq. (F1) decay doubly exponentially. Their largest magnitude is achieved for the $\mathbf{k}$ of smallest magnitude, which is $2\pi / L$, with $L = \max\{L_x, L_y, L_z\}$. With this observation, we can build an upper bound for the remainder series. Note that

$$\begin{aligned}\sum_{j=N}^{\infty} b^{2j} e^{-b^{2j}|\mathbf{k}|^2 \sigma^2/2} &= b^{2N} \sum_{j'=0}^{\infty} b^{2j'} e^{-b^{2j'}|\mathbf{k}|^2 \sigma^2/2} \left( e^{-(b^{2N}-1)b^{2j'}|\mathbf{k}|^2 \sigma^2/2} \right) \\ &\leq b^{2N} e^{-(b^{2N}-1) 2\pi^2 \sigma^2/L^2} \sum_{j'=0}^{\infty} b^{2j'} e^{-b^{2j'}|\mathbf{k}|^2 \sigma^2/2}\end{aligned}, \tag{F2}$$

where the inequality follows from $|\mathbf{k}| \geq 2\pi / L$ and $j' \geq 0$. Comparing the result to Eq. (F1), we infer that the absolute error for the far part of the periodic energy is bounded by

$$b^{2N} e^{(1-b^{2N}) 2\pi^2 \sigma^2/L^2} \left( 2\pi \ln(b) \sigma^2 V \sum_{\mathbf{k} \neq 0} |C_\mathbf{k}|^2 \sum_{j=0}^{\infty} b^{2j} e^{-b^{2j}|\mathbf{k}|^2 \sigma^2/2} \right). \tag{F3}$$

The expression in parentheses is the actual far part of the periodic energy. The relative error for the far part of the periodic energy is thus bounded by

$$\exp\left(2N\ln(b) - 2(b^{2N} - 1)\frac{\pi^2\sigma^2}{L^2}\right). \tag{F4}$$

**References**

1. P. Gibbon and G. Sutmann. Long-range interactions in many-particle simulation. *Quantum Simulations of Complex Many-Body Systems: From Theory to Algorithms*, lecture notes, eds. J. Grotendorst, D. Marx, and A. Muramatsu, 467–506 (John von Neummann Institute for Computing, Jülich, Germany, 2002).

2. C. Sagui and T. A. Darden. Molecular dynamics simulations of biomolecules: long-range electrostatic effects. *Annu. Rev. Biophys. Biomol. Struct.* **28**, 155–179 (1999).

3. M. Karttunen, J. Rottler, I. Vattulainen, and C. Sagui. Electrostatics in biomolecular simulations: Where are we now and where are we heading? *Computational Modeling of Membrane Bilayers*, ed. S. E. Feller. *Curr. Top. Membranes* **60**, 49–89 (Elsevier, New York, 2008).

4. A. D. MacKerell Jr. Empirical force fields for biological macromolecules: Overview and issues. *J. Comput. Chem.* **25(13)**, 1584–1604 (2004).

5. D. E. Shaw. A fast, scalable method for the parallel evaluation of distance-limited pairwise particle interactions. *J. Comput. Chem.* **26(13)**, 1318–1328 (2005).

6. Y. Sun, G. Zheng, C. Mei, E. J. Bohm, J. C. Phillips, L. V. Kalé, and T. R. Jones. Optimizing fine-grained communication in a biomolecular simulation application on Cray XK6. *Proceedings of the Conference on High-Performance Computing, Networking, Storage, and Analysis (SC12)* (IEEE, New York, 2012).

7. C. Kutzner, R. Apostolov, B. Hess, and H. Grubmüller. Scaling of the GROMACS 4.6 molecular dynamics code on SuperMUC. *Parallel Computing: Accelerating Computational Science and Engineering (CSE)*, eds. M. Bader, A. Bode, H.-J. Bungartz, M. Gerndt, G. R. Joubert, and F. Peters. *Adv. Parallel Comput.* **25**, 722–730 (IOS Press, Amsterdam, 2014).

8. P. P. Ewald. Die Berechnung optischer und elektrostatischer Gitterpotentiale. *Ann. Phys.* **369(3)**, 253–287 (1921).

9. J. W. Perram, H. G. Petersen, and S. W. de Leeuw. An algorithm for the simulation of condensed matter which grows as the 3/2 power of the number of particles. *Mol. Phys.* **65(4)**, 875–893 (1988).

10. V. Natoli and D. M. Ceperley. An optimized method for treating long-range potentials. *J. Comput. Phys.* **117**, 171–178 (1995).

11. T. Darden, D. York, and L. Pedersen. Particle mesh Ewald: An N·log(N) method for Ewald. *J. Chem. Phys.* **98(12)**, 10089–10092 (1993).

12. U. Essmann, L. Perera, M. L. Berkowitz. A smooth particle mesh Ewald method. *J. Chem. Phys.* **103(19)**, 8577–8593 (1995).

13. R. W. Hockney and J. W. Eastwood. *Computer Simulation Using Particles* (McGraw-Hill, New York, 1981).

14. D. York and W. Yang. The fast Fourier Poisson method for calculating Ewald sums. *J. Chem. Phys.* **101(4)**, 3298–3300 (1994).

15. Y. Shan, J. L. Klepeis, M. P. Eastwood, R. O. Dror, and D. E. Shaw. Gaussian split Ewald: A fast Ewald mesh method for molecular simulation. *J. Chem. Phys.* **122(5)**, 54101 (2005).

16. M. Levitt, M. Hirshberg, R. Sharon, and V. Daggett. Potential energy function and parameters for simulations of the molecular dynamics of proteins and nucleic acids in solution. *Comput. Phys. Comm.* **91(1–3)**, 215–231 (1995).

17. D. van der Spoel and P. J. van Maaren. The origin of layer structure artifacts in simulations of liquid water. *J. Chem. Theory Comput.* **2(1)**, 1–11 (2006).

18. E. F. Bertaut. Electrostatic potentials, fields and field gradients. *J. Phys. Chem. Solids* **39(2)**, 97–102 (1978).

19. D. M. Heyes. Electrostatic potentials and fields in infinite point charge lattices. *J. Chem. Phys.* **74(3)**, 1924 (1981).

20. P. F. Batcho and T. Schlick. New splitting formulations for lattice summations. *J. Chem. Phys.* **115(18)**, 8312–8326 (2001).

21. G. Beylkin and L. Monzón. On approximation of functions by exponential sums. *Appl. Comput. Harmon. Anal.* **19(1)**, 17–48 (2005).

22. G. Beylkin and L. Monzón. Approximation by exponential sums revisited. *Appl. Comput. Harmon. Anal.* **28(2)**, 131–149 (2010).

23. S. W. Smith. *The Scientist and Engineer's Guide to Digital Signal Processing* (California Technical Publishing, San Diego, 1997).

24. S. W. de Leeuw, J. W. Perram, and E. R. Smith. Simulation of electrostatic systems in periodic boundary conditions. I. Lattice sums and dielectric constants. *Proc. R. Soc. Lond. A* **373(1752)**, 27–56 (1980).

25. R. Harrison, G. Fann, T. Yanai, Z. Gan, and G. Beylkin. Multiresolution quantum chemistry: Basic theory and initial applications. *J. Chem. Phys.* **121(23)**, 11587–11598 (2004).

26. W. Kutzelnigg. Theory of the expansion of wave functions in a Gaussian basis. *Int. J. Quantum Chem.* **51(6)**, 447–462 (1994).

27. T. Laino, F. Mohamed, A. Laio, and M. Parrinello. An efficient real space multigrid QM/MM electrostatic coupling. *J. Chem. Theory Comput.* **1(6)**, 1176–1184 (2005).

28. T. Laino, F. Mohamed, A. Laio, and M. Parrinello. An efficient linear-scaling electrostatic coupling for treating periodic boundary conditions in QM/MM simulations. *J. Chem. Theory Comput.* **2(5)**, 1370–1378 (2006).

29. J. B. Klauda, R. M. Venable, J. A. Freites, J. W. O'Connor, D. J. Tobias, C. Mondragon-Ramirez, I. Vorobyov, A. D. MacKerell Jr., and R. W. Pastor. Update of the CHARMM all-atom additive force field for lipids: validation on six lipid types. *J. Phys. Chem. B* **114(23)**, 7830–7843 (2010).

30. M. Patra, M. Karttunen, M. T. Hyvönen, E. Falck, P. Lindqvist, and I. Vattulainen. Molecular dynamics simulations of lipid bilayers: Major artifacts due to truncating electrostatic interactions. *Biophys. J.* **84(6)**, 3636–3645 (2003).

31. S. E. Feller, R. W. Pastor, A. Rojnuckarin, S. Bogusz, and B. R. Brooks. Effect of electrostatic force truncation on interfacial and transport properties of water. *J. Phys. Chem.* **100(42)**, 17011–17020 (1996).

32. A. Aksimentiev and K. Schulten. Imaging alpha-hemolysin with molecular dynamics: Ionic conductance, osmotic permeability, and the electrostatic potential map. *Biophys. J.* **88(6)**, 3745–3761 (2005).

33. R. D. Skeel, I. Tezcan, and D. J. Hardy. Multiple grid methods for classical molecular dynamics. *J. Comput. Chem.* **23(6)**, 673–684 (2002).

34. J. E. Stone, J. C. Phillips, P. L. Freddolino, D. J. Hardy, L. G. Trabuco, and K. Schulten. Accelerating molecular modeling applications with graphics processors. *J. Comput. Chem.* **28(16)**, 2618–2640 (2007).

35. D. J. Hardy, J. E. Stone, and K. Schulten. Multilevel summation of electrostatic potentials using graphics processing units. *Parallel Comput.* **35(3)**, 164–177 (2009).
39

36. D. E. Shaw, J.P. Grossman, J. A. Bank, B. Batson, J. A. Butts, J. C. Chao, M. M. Deneroff, R. O. Dror, A. Even, C. H. Fenton, A. Forte, J. Gagliardo, G. Gill, B. Greskamp, C. R. Ho, D. J. Ierardi, L. Iserovich, J. S. Kuskin, R. H. Larson, T. Layman, L.-S. Lee, A. K. Lerer, C. Li, D. Killebrew, K. M. Mackenzie, S. Y.-H. Mok, M. A. Moraes, R. Mueller, L. J. Nociolo, J. L. Peticolas, T. Quan, D. Ramot, J. K. Salmon, D. P. Scarpazza, U. B. Schafer, N. Siddique, C. W. Snyder, J. Spengler, P. T. P. Tang, M. Theobald, H. Toma, B. Towles, B. Vitale, S. C. Wang, and C. Young. Anton 2: Raising the bar for performance and programmability in a special-purpose molecular dynamics supercomputer. *Proceedings of the Conference on High-Performance Computing, Networking, Storage, and Analysis (SC14)* (IEEE, New York, 2014).

37. L. N. Trefethen. *Finite Difference and Spectral Methods for Ordinary and Partial Differential Equations* (unpublished text, available at http://people.maths.ox.ac.uk/trefethen/pdetext.html, 1996).

**Figures and Tables**

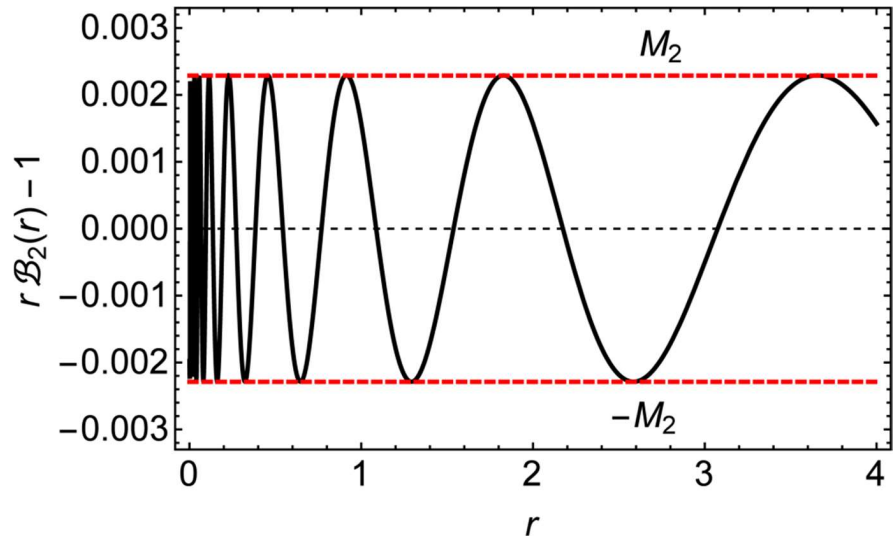

**Figure 1.** Relative error for the BSA with $b = 2$. Also plotted (dashed red lines) are the lower and upper bounds for the relative error. The magnitude of the bounds is independent of $\sigma$, which for this plot is equal to 1.

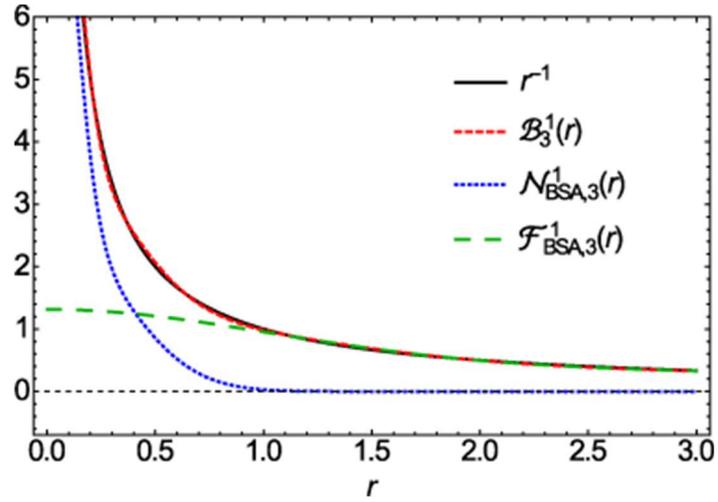

**Figure 2.** The near $\mathcal{N}^{\sigma}_{\text{BSA},b}(r)$ and far $\mathcal{F}^{\sigma}_{\text{BSA},b}(r)$ parts of the BSA series $\mathcal{B}^{\sigma}_{b}(r)$ approximating $1/r$, with $b = 3$ and $\sigma = 1$. With this large base $b$, the difference between $r^{-1}$ (black, solid line) and $\mathcal{B}^{\sigma}_{b}(r)$ (red, dashed line) is (just) visible. The near component (blue, dotted line) decays rapidly in real space, and the far component (green, dashed line) is smooth.

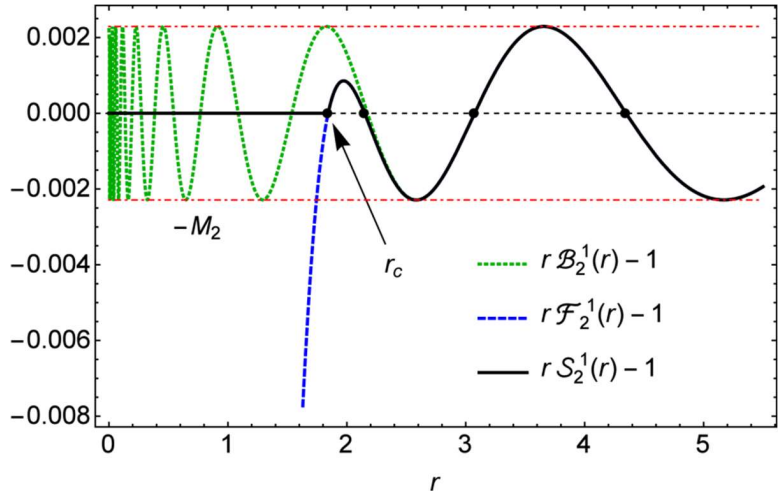

**Figure 3.** Relative error in approximating $1/r$ for the BSA (green, dotted line), its far part alone (blue, dashed line), and the $u$-series (black, solid line), all for $b = 2$ and $\sigma = 1$. The first four points where the BSA far part equals $1/r$ are marked with dots, and the first of these (which is equal to $r_0$ since $\sigma = 1$) is selected as the cutoff radius $r_c$. The $u$-series is defined to coincide with the BSA far part for $r \geq r_c$, so their relative errors also coincide for $r \geq r_c$. To the left of $r_c$, the $u$-series is constructed to be exactly $1/r$; its relative error on this domain is thus 0.

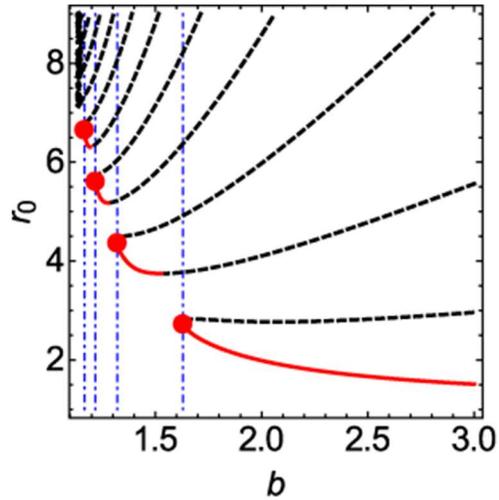

**Figure 4.** Roots of the relative error of the *u*-series far part given in Eq. (24), with $w_0$ chosen by way of Eq. (25). Each pair $(b, r_0)$ on the lower envelope of the family of curves corresponds to a $C^1$-continuous *u*-series. The solid red curves designate the set of optimal solutions that give rise to useful decompositions. This set is divided into countably many disjoint intervals (of which 4 are plotted). These intervals start at points (red bullets) for which the decomposition is $C^2$ continuous.

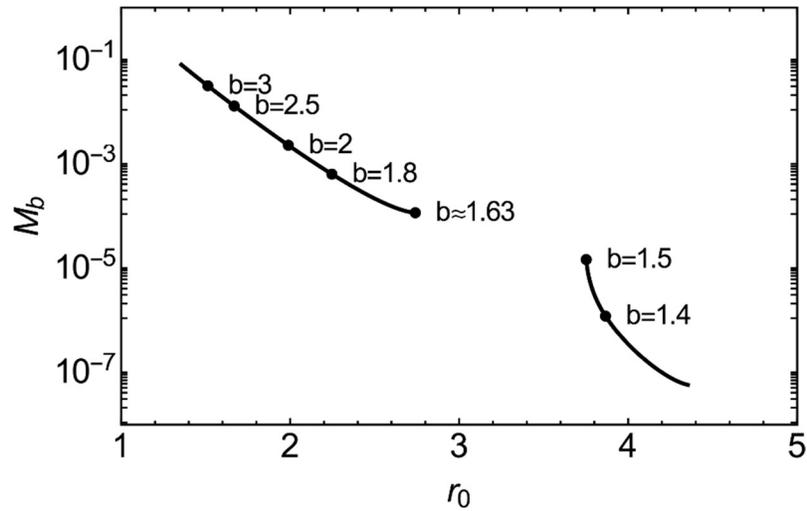

**Figure 5.** Relative error bound $M_b$ plotted against the smallest root $r_0$ of Eq. (26). Each value of $r_0$ implies a value of $b$, some of which are marked in the plot with bullets, including the $C^2$ solution at $b \approx 1.63$. The value of $b$ in turn defines the bound $M_b$ through Eq. (14). Within the first interval, the error decays roughly exponentially with $r_0$. The horizontal gap between the first and second intervals indicates a range of $r_0$ for which there is no solution for $b$, and is the same as the lowest vertical gap in Fig. 4.

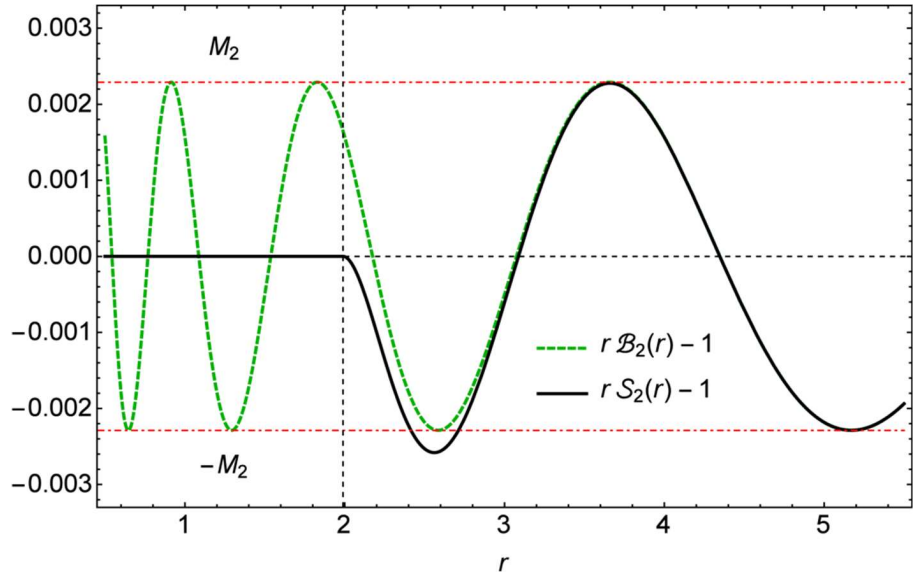

**Figure 6.** Relative errors for the BSA and the *u*-series $C^1$ construction for $b = 2$ and $\sigma = 1$. This *u*-series construction is $C^1$ continuous and slightly undershoots the BSA error to the right of $r_c = r_0 \approx 1.989$ before taking on the common asymptotic values for larger *r*. By comparison, the $C^0$-continuous *u*-series shown previously in Fig. 3 stays in bounds, but has a kink at $r_0 \approx 1.843$.

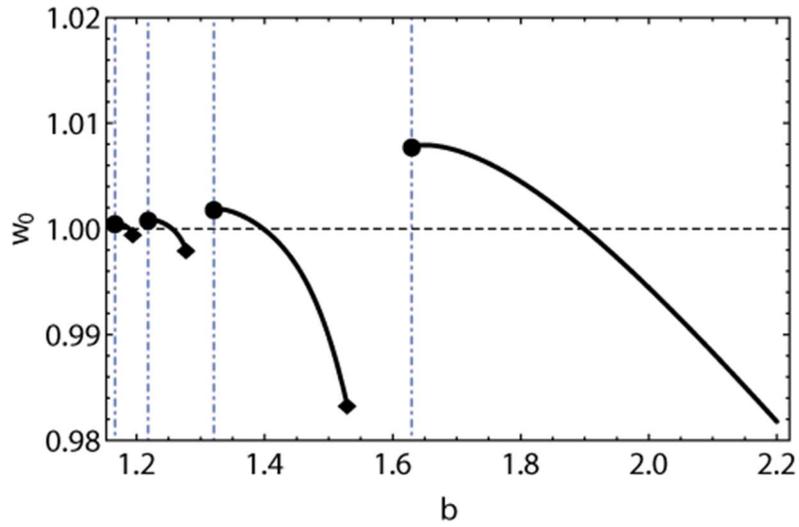

**Figure 7.** Coefficients $w_0$ of the narrowest Gaussian for the $C^1$ construction of the $u$-series. The coefficients converge to 1 as the base $b$ decreases toward 1, and do not deviate from 1 by more than 2% for the range $b \leq 2$, which has the most relevance in practice.

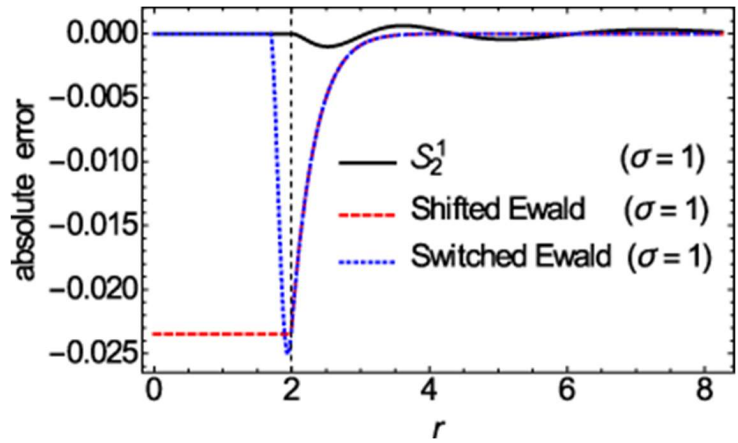

**Figure 8.** Absolute error in the interparticle potential for the Ewald decomposition (dashed red and blue lines) and the *u*-series (solid black line) with $b = 2$, for $\sigma = 1$ and $r_c = r_0 \approx 1.989$. The red Ewald line resolves the discontinuity at $r = r_c$ by shifting the potential, whereas the blue line uses a switching function over the range $(0.85 r_c, r_c)$.

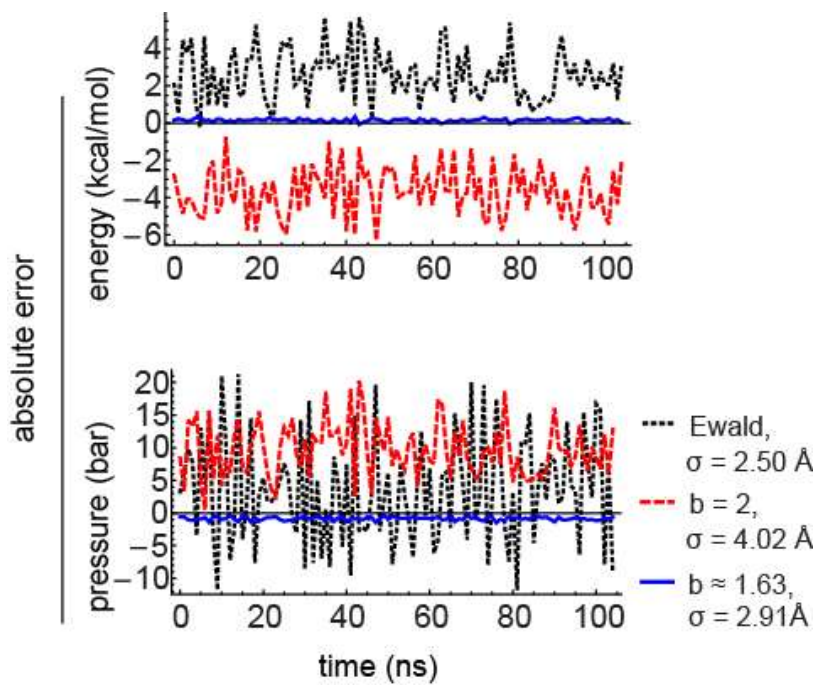

**Figure 9.** Errors in the energy and pressure with different electrostatic decompositions for 100 configurations of a DPPC system. The configurations were taken at 1-ns intervals from a single 100-ns simulation. All calculations used the cutoff radius $r_c = 8$ Å for the near electrostatics. A grid-based method with a very fine grid was used to evaluate the far part, so as not to introduce additional errors beyond those inherent to the decomposition scheme chosen. In red are the results of the $C^1$-continuous $u$-series with b = 2 and $\sigma = 4.02$ Å (the same ratio of $r_c / \sigma$ as used in Fig. 8). In black are the results of the Ewald decomposition with $\sigma = 2.50$ Å. The errors of the two decompositions are comparable in magnitude, despite the smaller value of $\sigma$ used for the Ewald decomposition. Also shown, in blue, are results of a simulation using the $u$-series with $b \approx 1.63$, corresponding to the first $C^2$-continuous solution from Table 1; even though this simulation uses $\sigma = 2.91$ Å, a larger value than in the simulations using the Ewald decomposition, it produces substantially more accurate results.

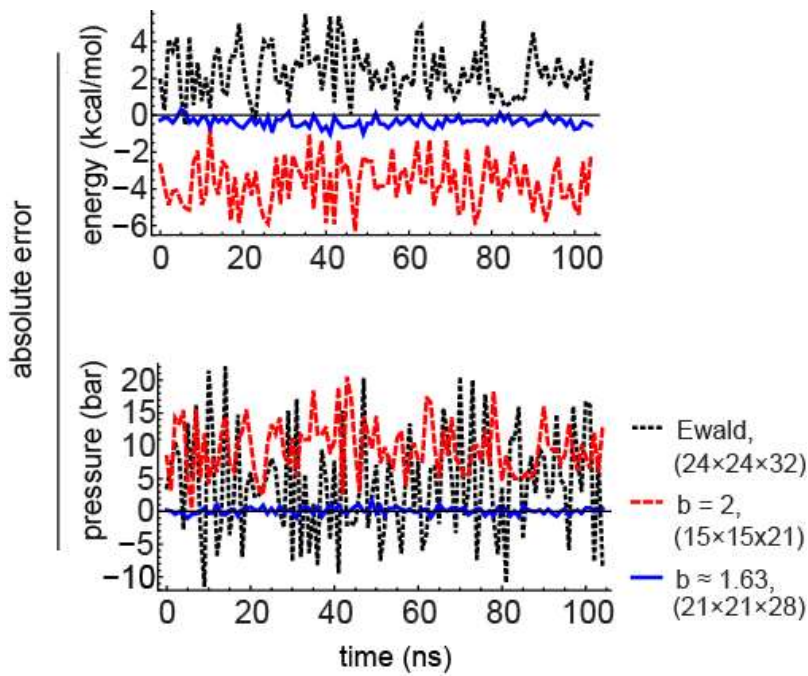

**Figure 10.** Errors in the energy and pressure for the same configurations of DPPC and same decomposition schemes as shown in Fig. 9, but with the far part evaluated using approximate methods typical of current simulation practice. Grid sizes used are given in parentheses in the legend. The larger value of $\sigma$ provided by the *u*-series allows coarser grids to be used. For Ewald and *u*-series with $b = 2$, the contribution to the error from the *k*-space sum is small, and the plots mirror those of Fig. 9. For the more accurate *u*-series with $b \approx 1.63$, the error from the *k*-space sum contributes more substantially to the total error.

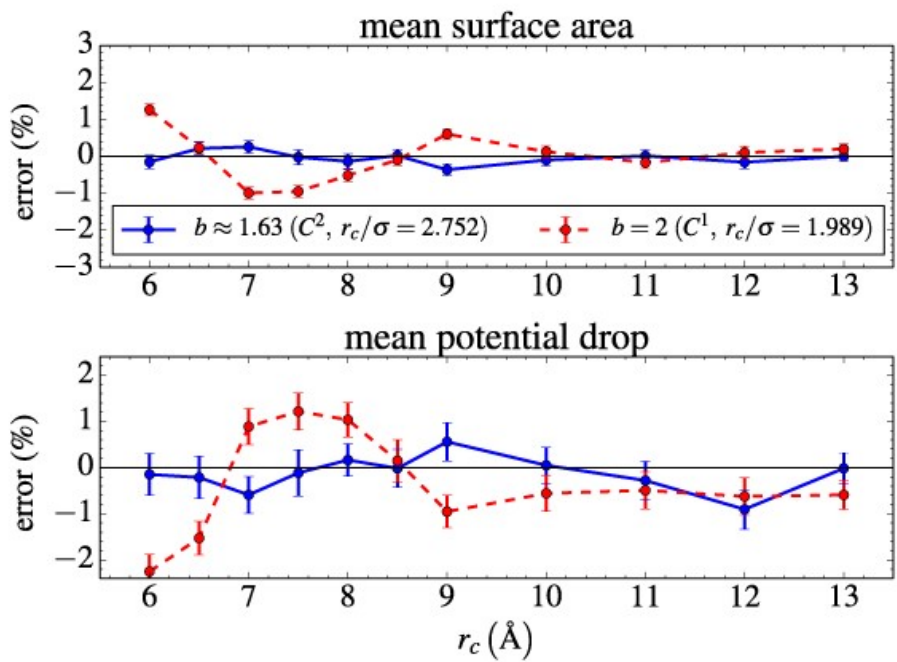

**Figure 11.** Errors in mean membrane properties as a function of the electrostatic cutoff radius $r_c$. The baseline simulation used Ewald with $r_c = 13.0$ Å, and error bars represent one standard error. The ranges of the y-axes are chosen to show the statistical fluctuations (± one standard deviation) of the measured property within the baseline run. The dashed red line is for the $u$-series with $b = 2$, whereas the solid blue line is for the more accurate $u$-series with $b \approx 1.63$.

**Table 1.** Parameters for the first three $C^2$-continuous solutions arising from the $C^1$ construction defined by Eq. (24), along with their relative error bound $M_b$.

| $b$ | $r_0$ | $w_0$ | $M_b$ |
|---|---|---|---|
| 1.62976708826776469 | 2.75200266680234223 | 1.00780697934380681 | $1.158 \times 10^{-4}$ |
| 1.32070036405934420 | 4.39145547116383425 | 1.00188914114811980 | $5.583 \times 10^{-8}$ |
| 1.21812525709410644 | 5.63552881512711037 | 1.00090146156033341 | $3.889 \times 10^{-11}$ |

**Table 2.** Parameters for the $C^0$, $C^1$, and $C^2$ constructions with $b = 2$ ($M_b = 2.289 \times 10^{-3}$). Refer to Eqs. (19), (24), and (28), respectively. The second row ($C^1$ construction with $b = 2$) is our recommended parameterization for most MD simulations.

| Continuity | $r_0$ | $w_0$ | $s_0$ |
|---|---|---|---|
| $C^0$ | 1.84309078551204634 | 1.00000000000000000 | 1.00000000000000000 |
| $C^1$ | 1.98925368390802627 | 0.994446492762232252 | 1.00000000000000000 |
| $C^2$ | 2.39427106444012661 | 0.892857784448501290 | 1.02474173246377476 |

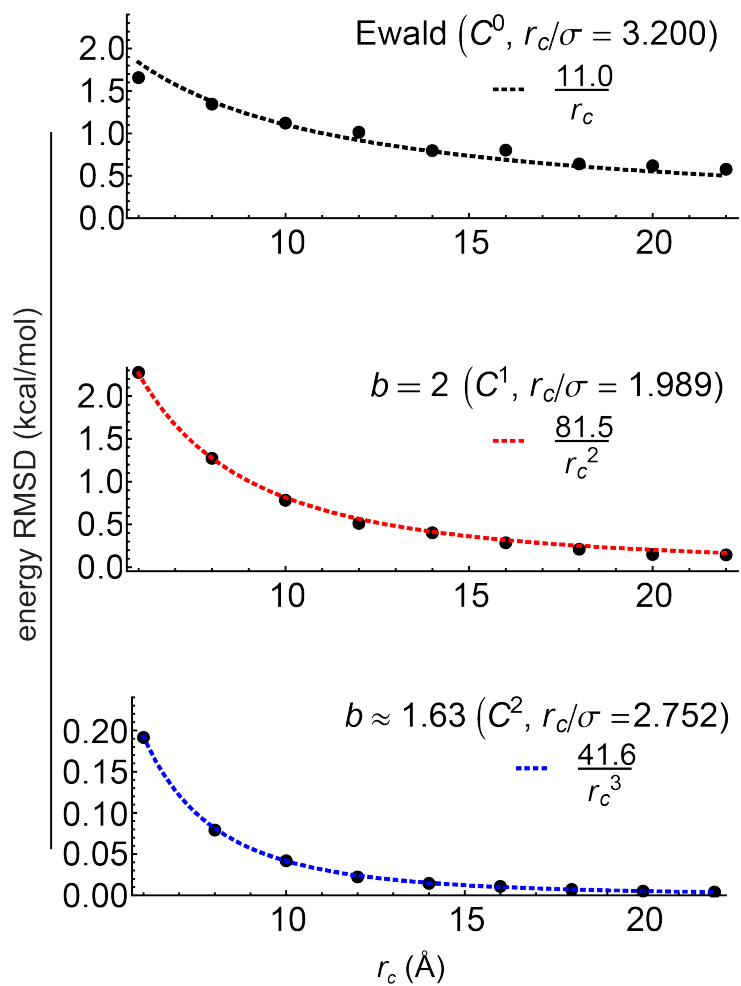

**Figure 12.** Root-mean-square deviations (RMSDs) of the errors for the electrostatic energies as functions of the electrostatic cutoff radius $r_c$. The dotted lines are best fits of the form $a / r^{d+1}$, with $a$ a real-number parameter and $d$ an integer parameter. The radii $r_c$ are varied such that the ratios $r_c / \sigma$ for the three methods are constant and equal to the ratios used for Fig. 9. The decay of the fluctuations of the energy errors appear to follow inverse polynomial laws, with the smoother approximations featuring faster asymptotic decay.

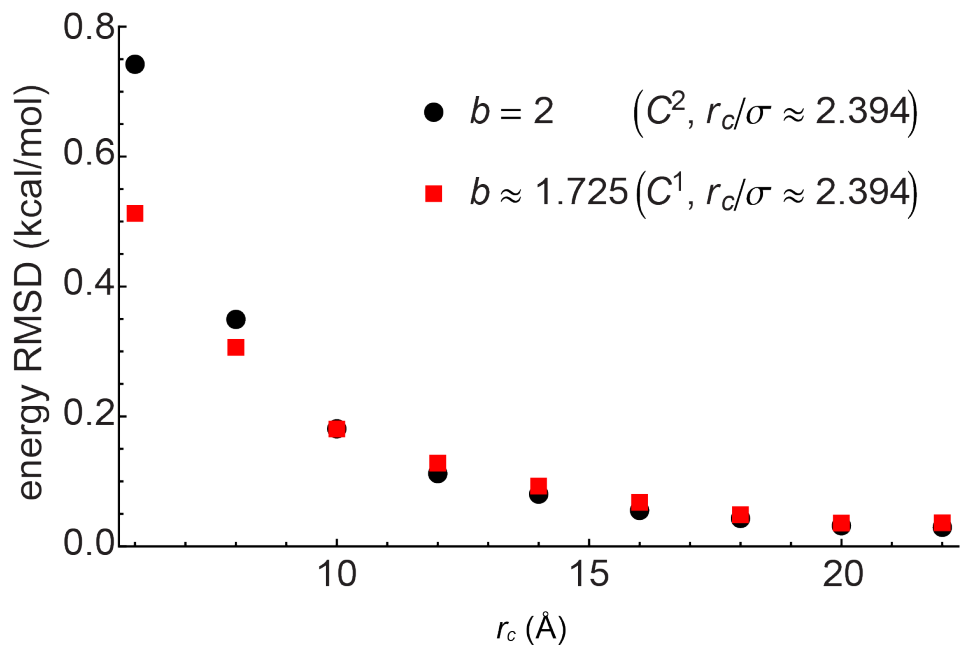

**Figure 13.** RMSDs of energy errors for two *u*-series sharing the same set of parameters ($r_c$, $r_c / \sigma$), but different levels of smoothness. Although the smoother *u*-series is expected to have smaller errors for very large $r_c$, it is not a decidedly superior method for the range of cutoff radii displayed.